\documentclass[12pt]{article}
\begin{document}

\title{{\bf Generalized Jarzynski Equality}
\thanks{Alberta-Thy-11-12, arXiv:yymm.nnnn [hep-th]}}
\author{
Don N. Page
\thanks{Internet address:
profdonpage@gmail.com}
\\
Department of Physics\\
4-181 CCIS\\
University of Alberta\\
Edmonton, Alberta T6G 2E1\\
Canada
}

\date{2012 July 13}

\maketitle
\large
\begin{abstract}
\baselineskip 25 pt

The Jarzynski equality equates the mean of the exponential of the negative of the work (per fixed temperature) done by a changing Hamiltonian on a system, initially in thermal equilibrium at that temperature, to the ratio of the final to the initial equilibrium partition functions of the system at that fixed temperature.  It thus relates two thermal equilibrium quantum states.  Here a generalization is given that relates any two quantum states of a system.

\end{abstract}

\normalsize

\baselineskip 23 pt

\newpage

For a quantum system with an initial Hamiltonian $H_0$ that gives energy eigenstates $|i\rangle$ with energy eigenvalues $E_i$ that lead to positive heat
capacity, at fixed inverse temperature $\beta$ = $1/(kT)$ one may define the initial partition function
\begin{equation}
Z_0 \equiv tr\left(e^{-\beta H_0}\right) = \sum_i e^{-\beta E_i}.
\label{partition}
\end{equation}
The (mixed) thermal equilibrium state or Gibbs state (the canonical ensemble) for this initial Hamiltonian at this temperature is
\begin{equation}
\rho_0 = Z_0^{-1} tr\left(e^{-\beta H_0}\right)
       \equiv \sum_i p_i|i\rangle\langle i|,
\label{Gibbs}
\end{equation}
with the probability for the energy eigenstate $|i\rangle$ being
\begin{equation}
p_i = Z_0^{-1} e^{-\beta E_i}.
\label{prob}
\end{equation}

Differentiating the partition function $Z_0$ with respect to $\beta$ while keeping the initial Hamiltonian $H_0$ fixed gives
\begin{equation}
\frac{\partial Z_0}{\partial\beta} = -tr\left(H_0 e^{-\beta H_0}\right)
= -\sum_i E_i e^{-\beta E_i} = -Z_0\sum_i p_i E_i,
\label{diff}
\end{equation}
so the expectation value of the initial energy is
\begin{equation}
\langle E_0 \rangle \equiv tr\left(H_0\rho_0\right)
= \sum_i p_i E_i = - \frac{\partial\ln{Z_0}}{\partial\beta}.
\label{energy}
\end{equation}

Furthermore, the initial equilibrium von Neumann entropy is
\begin{equation}
S_0 \equiv -tr\left(\rho_0\ln{\rho_0}\right)
  = -\sum_i p_i\ln{p_i}
  = -\sum_i p_i(-\ln{Z_0}-\beta E_i)
  = \ln{Z_0} + \beta\langle E_0 \rangle.
\label{entropy}
\end{equation}
The initial equilibrium Helmholtz free energy is then
\begin{equation}
F_0 \equiv \langle E_0 \rangle - kT S_0 = -kT\ln{Z_0} = -(1/\beta)\ln{Z_0}.
\label{free}
\end{equation}

Now suppose the system is initially in the thermal equilibrium mixed state (Gibbs canonical ensemble) at the given temperature $T$ or inverse temperature $\beta$, but then over some period of time the Hamiltonian changes from its initial form $H_0$ to a final form $H_1$ with energy eigenstates $|j\rangle$ having energy eigenvalues $E_j$.  Here I shall assume that the system is a closed quantum system during this process, not able to exchange heat with any heat bath.  Suppose that replacing the subscripts 0 by 1 in the formulae above gives the final equilibrium partition function $Z_1$, final Gibbs state $\rho_1$, final equilibrium energy expectation value $\langle E_1 \rangle$, final equilibrium entropy $S_1$, and final Helmholtz free energy $F_1$.  Note that these are hypothetical final thermal equilibrium values, and {\it not} the actual values that one would get from the evolution of the system while the Hamiltonian is changed from $H_0$ to $H_1$.  (For example, since I am assuming that the system is a closed quantum system, the final von Neumann entropy would be the same as the initial entropy $S_0$.)

In the Heisenberg picture, let
\begin{equation}
U_{ji} = \langle j|i \rangle
\label{transamp}
\end{equation}
be the transition amplitude from the initial Hamiltonian $H_0$ energy eigenstate $|i\rangle$ (after its evolution by the changing Hamiltonian) to a final energy eigenstate $|j\rangle$ of the final Hamiltonian $H_1$, so that the actual final mixed state (not the equilibrium Gibbs state for the final Hamiltonian) is
\begin{equation}
\rho = \sum_{i,j,j'} p_i U_{ji}U_{ij'}^\dagger |j\rangle\langle j'|.
\label{finalstate}
\end{equation}

Given an initial energy eigenstate $|i\rangle$, the transition probability that it becomes the final energy eigenstate $|j\rangle$ is then
\begin{equation}
P_{ji} = |\langle j|i \rangle|^2 = U_{ji}U_{ij}^\dagger.
\label{transprob}
\end{equation}
One can readily see that the unitarity of the transition matrix gives
\begin{equation}
\sum_i P_{ji} = \sum_i U_{ji}U_{ij}^\dagger = \delta_{jj} = 1,
\label{sumrule}
\end{equation}
so that the sum of the transition probabilities from all the initial states $|i\rangle$ to any particular final state $|j\rangle$ is unity.

Since the probability of starting in the initial eigenstate $|i \rangle$ is $p_i = Z_0^{-1}e^{-\beta E_i}$, the joint probability to start in the initial energy eigenstate $|i\rangle$ and to end up in the final energy eigenstate $|j\rangle$ is
\begin{equation}
P_{i\& j} = p_i P_{ji} = Z_0^{-1}e^{-\beta E_i} P_{ji}.
\label{jointprob}
\end{equation}

The work done on the system by the changing Hamiltonian if the energy changes from $E_i$ initially to $E_j$ finally is
\begin{equation}
W_{ij} = E_j - E_i.
\label{work}
\end{equation}
For any function of this work during the process of changing the Hamiltonian, say $f(W_{ij})$, we can define the mean as
\begin{equation}
\overline{f(W_{ij})} \equiv \sum_{i,j} P_{i\& j} f(W_{ij}).
\label{mean}
\end{equation}

Now Jarzynski \cite{J1,J2,J3,J4,J5} has given the following equality (here in the case of a process in which the system is not in contact with a heat bath):
\begin{equation}
\overline{e^{-\beta W_{ij}}} = e^{-\beta(F_1-F_0)} \equiv Z_1/Z_0.
\label{Jarzynski}
\end{equation}
This Jarzynski equality can be easily proved from the definitions given above as follows:
\begin{eqnarray}
\overline{e^{-\beta W_{ij}}}
&\equiv& \sum_{i,j} P_{i\& j} e^{-\beta W_{ij}} \nonumber \\
&=& \sum_{i,j} Z_0^{-1}e^{-\beta E_i} P_{ji} e^{-\beta (E_j - E_i)} \nonumber \\
&=& \sum_{i,j} Z_0^{-1}P_{ji} e^{-\beta E_j} \nonumber \\
&=& Z_0^{-1}\sum_j e^{-\beta E_j}\sum_i P_{ji} \nonumber \\
&=& Z_0^{-1}Z_1(1) \nonumber \\
&=& Z_1/Z_0.
\label{Jarproof}
\end{eqnarray}

The point of this Letter is that there is nothing in this proof of the Jarzynski equality that requires $H_0$ and $H_1$ actually to be Hamiltonians, so long as the partition functions, free energies, energy eigenstates, transition probabilities, etc.\ are defined accordingly in terms of arbitrary Hermitian operators used in place of $H_0$ and $H_1$.  In particular, in place of the initial and final Gibbs states, one can use any mixed states
\begin{equation}
\rho_0 = \sum_i p_i|i\rangle\langle i|,\ \ \ \ 
\rho_1 = \sum_j q_j|j\rangle\langle j|,
\label{states}
\end{equation}
with eigenvalue sets $\{p_i\}$ and $\{q_j\}$ that each are nonnegative real numbers that add up to unity, and with the corresponding orthonormal eigenstate sets being $\{|i\rangle\}$ and $\{|j\rangle\}$.  One can also choose `partition function' values $Z_0$ and $Z_1$ to be arbitrary real positive numbers.  Then for any fixed temperature $T$ and $\beta = 1/(kT)$, one can define Hermitian operators with the dimensions of energy that are
\begin{eqnarray}
H_0 &\equiv&
     -kT \ln{(Z_0 \rho_0)} = -kT \sum_i (\ln{Z_0}+\ln{p_i})|i\rangle\langle i|,
\nonumber \\
H_1 &\equiv& 
     -kT \ln{(Z_1 \rho_1)} = -kT \sum_j (\ln{Z_1}+\ln{q_j})|j\rangle\langle j|,
\label{operators}
\end{eqnarray}
so that
\begin{equation}
\rho_0 = Z_0^{-1} tr\left(e^{-\beta H_0}\right),\ \ \ \ 
\rho_1 = Z_1^{-1} tr\left(e^{-\beta H_1}\right).
\label{simpleGibbs}
\end{equation}

Therefore, the eigenstates of $H_0$ are the eigenstates $|i\rangle$ of the mixed state $\rho_0$, and the corresponding eigenvalues of $H_0$ are $-kT(\ln{Z_0}+\ln{p_i})$.  Similarly, the eigenstates of $H_1$ are the eigenstates $|j\rangle$ of the mixed state $\rho_1$, and the corresponding eigenvalues of $H_1$ are $-kT(\ln{Z_1}+\ln{q_j})$.

If $\rho_0$ and/or $\rho_1$ have zero eigenvalues, $p_i = 0$ and/or $q_j = 0$, the resulting $H_0$ and/or $H_1$ will have infinite eigenvalues and hence give infinity when acting on any states that have nonzero amplitudes to be any of the eigenstates of zero eigenvalue of $\rho_0$ and/or $\rho_1$.  However, for the generalized Jarzynski equality, only the eigenstates of nonzero eigenvalues contribute, so there is no problem with infinities.

For fixed mixed states $\rho_0$ and $\rho_1$, it is simplest to choose $kT = Z_0 = Z_1 = 1$.  Then if we call the corresponding Hermitian operators $h_0$ and $h_1$, we get
\begin{eqnarray}
h_0&=&-\ln{\rho_0} = -\sum_i \ln{p_i}|i\rangle\langle i|,
\nonumber \\
h_1&=&-\ln{\rho_1} = -\sum_j \ln{q_j}|j\rangle\langle j|,
\label{simplifiedoperators}
\end{eqnarray}
with respective eigenvalues
\begin{equation}
e_i = -\ln{p_i},\ \ \ \ e_j = -\ln{q_j}.
\label{simplifiedstates}
\end{equation}
Then if $w_{ij} \equiv e_j - e_i$, defining
\begin{equation}
\overline{f(w_{ij})} \equiv \sum_{i,j} p_i |\langle j|i \rangle|^2 f(w_{ij})
\label{simplemean}
\end{equation}
leads to the generalized Jaryzynski equality
\begin{equation}
\overline{e^{-w_{ij}}} = 1.
\label{generalizedJarzynski}
\end{equation}

Because the ordinary Jarzynski equality refers to thermal equilibrium Gibbs states in canonical ensembles, which often do not exist in the presence of gravity, the generalized Jarzynski equality would be more applicable to gravitational systems.

I am grateful to Bei-Lok Hu for introducing me to the Jarzynski equality and for many discussions about it, and I thank Rafael Sorkin for helping me arrive at the simple proof for it given in Eq.\ (\ref{Jarproof}).  This research was supported in part by the Natural Sciences and Engineering Research Council of Canada.


\baselineskip 4pt

\begin{thebibliography}{99}

\bibitem{J1} C.~Jarzynski, Phys.\ Rev.\ Lett.\ {\bf 78}, 2690-2693 (1997).

\bibitem{J2} C.~Jarzynski, Phys.\ Rev.\ E{\bf 56}, 5018-5035 (1997).

\bibitem{J3} C.~Jarzynski, J.\ Stat.\ Phys.\ {\bf 98}, 77-102 (2000).

\bibitem{J4} C.~Jarzynski, Euro.\ Phys.\ J.\ B{\bf 64}, 331-340 (2008).

\bibitem{J5} C.~Jarzynski, Ann.\ Rev.\ Cond.\ Matt.\ Phys.\ {\bf 2}, 329-351 (2011).

\end{thebibliography}
\end{document}